\documentclass{article}
\usepackage{microtype}
\usepackage{graphicx}
\usepackage{tikz}
\usepackage{amsmath}
\usepackage{array}
\usepackage{booktabs}
\usepackage{tabularx}
\usepackage{listings}
\usepackage{subcaption}
\usepackage{dblfloatfix}
\usepackage{hyperref}

\usepackage[accepted]{mlsys2025}
\makeatletter
\renewcommand{\Notice@String}{}
\makeatother
\hypersetup{pdfsubject={},pdfkeywords={}}
\usepackage{circledsteps}
\usepackage{tikz}
\usepackage{xurl}
\usetikzlibrary{
    fit,
    arrows.meta,
    positioning,
    calc
}

\mlsystitlerunning{Chiron: Cache Management Empowered by LLM Intelligence for AI Workloads}

\newcommand{\chiron}{\textit{Chiron}}
\newenvironment{compactitems}{%
  \setlength{\leftmargini}{1.1em}%
  \begin{itemize}\setlength{\itemsep}{1pt}\setlength{\parskip}{0pt}}%
  {\end{itemize}}

\begin{document}

\twocolumn[
\mlsystitle{Chiron: Cache Management Empowered by LLM Intelligence for AI Workloads}

\begin{center}
\textbf{Bohan Zhang}\quad
\textbf{Renshu Ge}\quad
\textbf{Tianze Wang}\quad
\textbf{Chen Chen}\par\vspace{6pt}
Shanghai Jiao Tong University
\end{center}
\hypersetup{pdfauthor={Bohan Zhang, Renshu Ge, Tianze Wang, Chen Chen}}

\vskip 0.3in

\begin{abstract}
AI clusters keep data in remote object stores and rely on local caches to stop GPUs from waiting for data.
A cache performs well only when its prefetching and eviction policies match the access pattern, yet these patterns differ across AI workloads and change between phases of one workflow.
Existing methods either need code changes, which compromises user transparency, or use low-end heuristics, which cannot follow such changes.
Large language models (LLMs) can infer access behavior without user hints, but direct LLM control of cache actions suffers from long reasoning delay, high token cost, and poor decision quality.
We present \chiron{}, a cache system based on \emph{advisory switching control}.
An LLM asynchronously diagnoses the access pattern and recommends a policy.
A deterministic harness gates this advice, executes expert-designed policies locally, and returns cache feedback for review.
We implement \chiron{} atop JuiceFS and evaluate it on nine AI workloads.
It reduces job completion time by 14.3--41.2\% over the best baseline on popular AI workloads.
\end{abstract}
]

\section{Introduction}
\label{sec:introduction}

AI clusters run diverse workloads, from data preprocessing to model training and evaluation~\cite{resnet,lora,dreambooth,fno}.
As datasets and models grow, these clusters keep data in remote object stores such as Amazon S3~\cite{aws-s3} and serve it through a local cache layer such as Alluxio or JuiceFS~\cite{alluxio-cache,juicefs-cache}.
Remote accesses can stall input pipelines and leave GPUs waiting for data~\cite{data-stalls-dnn,tfdata,plumber}, while local capacity is limited.
Prefetching and eviction policies therefore crucially affect the cache hit ratio (CHR) and job completion time (JCT).

AI workloads make these policies hard to set.
Sequential, random, and skewed accesses are all common, each prefers different policies, and one workflow may pass through all three as it moves between phases (Section~\ref{sec:bg-workflow-diversity}).
An AI cache thus needs both \emph{adaptivity}, matching its policy to the current pattern, and \emph{transparency}, doing so without user hints (Section~\ref{sec:bg-cache-goals}).
Existing methods do not provide both (Section~\ref{sec:mot-workflow-adaptation}).
Code-intrusive adaptations such as Quiver~\cite{quiver} and NoPFS~\cite{nopfs} exploit workload knowledge through customized interfaces, which compromises transparency.
Low-end adaptations such as ARC~\cite{arc}, LeCaR~\cite{lecar}, and SFP~\cite{sfp} need no hints, but they follow variations within one pattern rather than transitions between patterns, which compromises efficacy.

Providing both requires inferring the upper-level computation logic from low-level accesses, a task that fits large language models (LLMs), which already assist networking~\cite{netllm} and database~\cite{dbot} management.
Directly controlling the cache with an LLM, however, is impractical.
We build a prototype in which an LLM lists the blocks to prefetch or evict after every read, and it reveals three challenges (Section~\ref{sec:mot-e2e-control}):
\begin{compactitems}
  \item \textbf{Long reasoning delay.} Qwen3.8 Flash takes 41\,s at P50 even with thinking off, while a storage client can issue hundreds of reads per second.
  \item \textbf{High token cost.} 128 reads consume over 10\,M tokens, so a workflow of about 60k reads would need about 5 billion.
  \item \textbf{Poor decision quality.} The model chooses among up to 1,963 candidate blocks and may hallucinate; the best prefetching accuracy across models is 48.2\% lower than that of SFP.
\end{compactitems}

We present \chiron{}, a fast, cheap, and effective cache system for AI workloads based on \emph{advisory switching control}.
As in supervisory switching control~\cite{morse1996supervisory,liberzon2003switching}, the LLM recognizes the operating regime and selects among expert-designed policies instead of acting on individual blocks.
Since an LLM may hallucinate, it only advises: a deterministic harness decides whether to adopt its advice and returns cache feedback (Section~\ref{sec:design}).
This design addresses the three challenges as follows:
\begin{compactitems}
  \item \textbf{Asynchronous diagnosis.} Inference stays off the read path: reads continue under the active policy while the LLM reasons in the background. This suffices because access patterns persist across many reads.
  \item \textbf{Pattern-level advice.} The LLM returns a short pattern label with a confidence score rather than block lists, and it runs only for routine checks or when cache feedback requests a review. One diagnosis then serves many reads, which keeps token cost low.
  \item \textbf{Harnessed execution.} Choosing among three patterns replaces choosing among thousands of blocks. Deterministic executors pick objects locally, a confidence gate defers uncertain advice, and feedback-triggered review revises outdated diagnoses, which contains hallucination.
\end{compactitems}

We implement \chiron{} atop JuiceFS with about 4,000 LoC of Go and 400 LoC of Python, and evaluate it on nine AI workloads with data on remote storage (Section~\ref{sec:evaluation}).
On three multi-phase workflows, \chiron{} achieves the highest CHR and reduces JCT by 14.3--41.2\% over the best baseline.
Ablations confirm that structured statistics, confidence gating, and feedback reflection are each necessary.
In a simulation with a 1\,TB dataset, \chiron{} also has the highest CHR and lowest JCT at every tested cache size.

\section{Background}
\label{sec:background}

\subsection{Caches Used for AI Workloads}
\label{sec:bg-workflows}

\begin{figure}[t]
\centering
\includegraphics[width=\columnwidth]{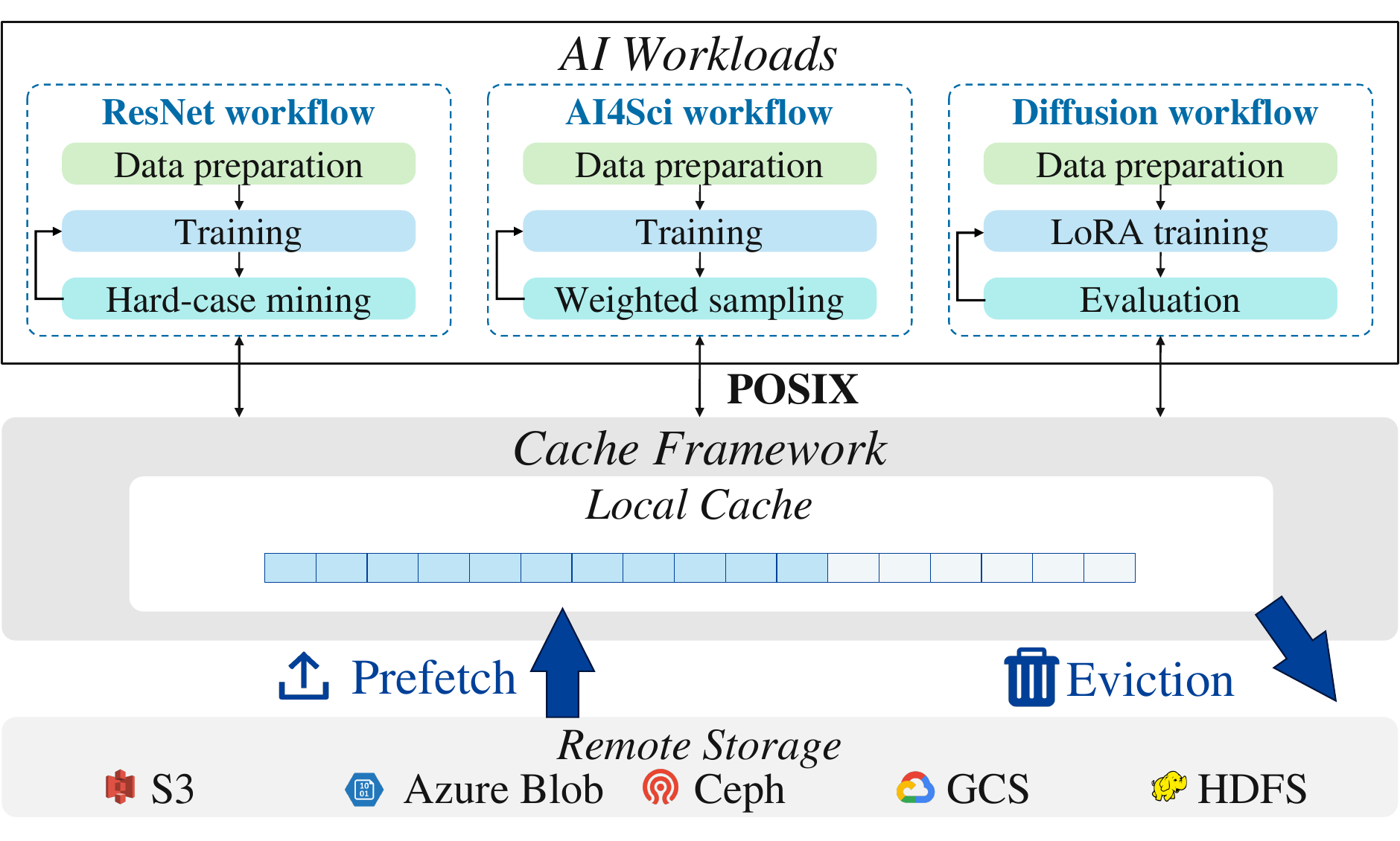}
\caption{Caching across AI workloads and workflow phases.}
\label{fig:bg-cache-workflows}
\end{figure}

AI techniques have substantially transformed many fields such as \emph{computer vision}~\cite{resnet}, \emph{natural language processing}~\cite{vaswani2017attention}, and autonomous driving~\cite{bojarski2016end}.
Developing them requires diverse workloads such as data preprocessing, training, and inference, which are hosted in GPU clusters.
With the drastic increase of dataset and model sizes, many enterprises such as Tencent~\cite{alluxio-tencent-game-ai} and RedNote~\cite{alluxio-rednote} now store data remotely in cloud object stores such as AWS S3~\cite{aws-s3} and Azure Blob~\cite{azure-blob-storage}.
Such a compute-storage disaggregated architecture (Figure~\ref{fig:bg-cache-workflows}) allows storage capacity to scale independently of GPU resources, attaining low cost and high reliability.

Under compute-storage disaggregation, caches are commonly adopted to improve remote I/O performance.
Accessing remote storage incurs network latency and consumes bandwidth, which can stall input pipelines and leave GPUs waiting for data~\cite{data-stalls-dnn,tfdata,plumber}; to mitigate such overhead, frameworks such as Alluxio and JuiceFS provide a caching layer between computation and remote storage~\cite{alluxio-cache,juicefs-cache}.
As shown in Figure~\ref{fig:bg-cache-workflows}, workflows access data through a POSIX-facing cache framework that retains data locally and serves cache hits without remote I/O.

Local cache capacity is limited relative to the datasets it serves, so appropriate cache-management policies are essential.
Two key aspects are \emph{prefetching}, which decides what data to fetch before it is requested~\cite{nopfs,garg2024crossprefetch,sfp}, and \emph{eviction}, which decides what resident data to remove when space is needed~\cite{qureshi2007adaptive,yang2023fifo,zhang2024sieve}.
These policies crucially affect the cache hit ratio and further the end-to-end execution time of workloads.

\subsection{Unique I/O Properties of AI Workloads}
\label{sec:bg-workflow-diversity}

\begin{table*}[t]
\centering
\caption{Representative file access patterns with example workloads, access sequences, and desired cache policies.}
\label{tab:access-patterns}
\scriptsize
\renewcommand{\arraystretch}{1.15}
\setlength{\tabcolsep}{3pt}
\begin{tabularx}{\textwidth}{@{}>{\raggedright\arraybackslash}p{0.08\textwidth}
                              >{\raggedright\arraybackslash}p{0.32\textwidth}
                              >{\raggedright\arraybackslash}X
                              >{\raggedright\arraybackslash}p{0.16\textwidth}
                              >{\raggedright\arraybackslash}p{0.14\textwidth}@{}}
\toprule
\textbf{Temporal patterns} & \textbf{Workloads} & \textbf{Example access sequence} &
\textbf{Desired prefetching policy} & \textbf{Desired eviction policy} \\
\midrule
Sequential & \begin{tabular}[t]{@{}l@{}}LLM model loading~\cite{serverlessllm},\\LLM tokenization~\cite{raffel2020t5}, $\cdots$\end{tabular} &
{\scriptsize\begin{tabular}[t]{@{}l@{}}\texttt{1.jpeg} $\to$ \texttt{2.jpeg} $\to$\\\texttt{3.jpeg} $\to$ \texttt{4.jpeg}\end{tabular}} &
Stride prefetching~\cite{chen1995effective} & Eager eviction~\cite{li2008tap} \\
\addlinespace[4pt]
Random & \begin{tabular}[t]{@{}l@{}}Vision fine-tuning~\cite{dosovitskiy2021vit},\\Speech fine-tuning~\cite{seo2021wav2kws}, $\cdots$\end{tabular} &
{\scriptsize\begin{tabular}[t]{@{}l@{}}\texttt{10969.jpeg} $\to$ \texttt{13625.jpeg} $\to$\\\texttt{19003.jpeg} $\to$ \texttt{8077.jpeg}\end{tabular}} &
No prefetching~\cite{azure_random} & No eviction~\cite{data-stalls-dnn} \\
\addlinespace[4pt]
Skewed & \begin{tabular}[t]{@{}l@{}}Recommendation training~\cite{dlrm},\\RAG retrieval~\cite{jin2025ragcache}, $\cdots$\end{tabular} &
{\scriptsize\begin{tabular}[t]{@{}l@{}}\texttt{A.csv} $\to$ \texttt{B.csv} $\to$\\\texttt{A.csv} $\to$ \texttt{A.csv}\end{tabular}} &
Neighborhood prefetching~\cite{li2004c} & LRU eviction~\cite{mattson-storage-hierarchies} \\
\bottomrule
\end{tabularx}
\end{table*}

Effective caching depends on how workloads traverse data, and AI workloads differ fundamentally from conventional ones~\cite{arc,gu2023adaptive,floratou2016adaptive} in two aspects.

\textit{(1) Diverse temporal access patterns across AI workloads.}
AI clusters support a wide range of workloads (Figure~\ref{fig:bg-cache-workflows}), including model training~\cite{qwen35-omni-report,deepseek-v4-report,kimi-k3-report}, model evaluation~\cite{lmms-eval,ueval,nvidia-aiperf}, and data preprocessing~\cite{fineweb2,flux-data}.
These tasks differ in how they traverse data (Table~\ref{tab:access-patterns}).
There are in general three typical patterns: \emph{sequential}, for workloads like preprocessing and testing; \emph{random}, for workloads like training~\cite{data-stalls-dnn}; and \emph{skewed} (i.e., with hotspots), for workloads like quality analysis and RAG~\cite{jin2025ragcache}.

\textit{(2) Dynamic temporal access patterns within one AI workflow.}
An AI development workflow may comprise multiple dependent tasks, each a \emph{phase} such as preprocessing, training, or evaluation (Figure~\ref{fig:bg-cache-workflows}).
For example, NVIDIA NeMo connects data curation, model training, evaluation, and deployment in one workflow~\cite{nvidia-llama4-inference}, and data-generation workflows alternate generation, filtering, de-duplication, and example selection~\cite{nvidia-financial-curation}.
The access pattern over one dataset may thus change at any moment and is hard to identify in one shot.

In summary, AI workloads exhibit high diversity and dynamicity in temporal I/O characteristics, involving three \emph{comparably common} access patterns.
By contrast, conventional workloads such as operating systems~\cite{denning1968working}, databases~\cite{zhou2025tiered}, and big data analytics~\cite{ananthanarayanan2012pacman} usually have one \emph{dominant} pattern, typically skewed.
This distinctness yields distinct design requirements for AI caches.

\subsection{Design Requirements of AI Caches}
\label{sec:bg-cache-goals}

\textbf{Caching preferences follow I/O patterns.}
Different access patterns desire different prefetching and eviction policies (Table~\ref{tab:access-patterns}).
Sequential accesses prefer \emph{stride prefetching} (i.e., prefetching sequential items) and \emph{eager eviction} (i.e., removing an item once it is accessed).
Random accesses prefer \emph{no prefetching}, as a low success rate wastes bandwidth and can pollute the cache~\cite{azure_random}, and \emph{no eviction}, as eviction has no impact when each sample is accessed once per epoch~\cite{data-stalls-dnn}.
Skewed accesses prefer \emph{neighborhood prefetching} (e.g., based on data-path locality) and \emph{LRU eviction}, the classical policy adopted in caching frameworks like Alluxio.
A mismatched policy lowers the cache hit ratio and prolongs job completion (Section~\ref{sec:evaluation}).

\textbf{Cache management requirements: adaptivity and transparency.}
An AI developer usually maintains a dedicated cache~\cite{juicefs_dedicated,alluxio_dedicated}, so the policy should match each workload at each instant.
This adaptivity must not come at the price of user transparency: application hints through customized cache APIs bring a code-intrusion burden that is impractical for public users~\cite{quiver}.
It is thus desirable for the cache to select suitable policies automatically, without user hints.
In a nutshell, we need an easy-to-use (with preserved user transparency), high-efficacy (with good workload adaptivity) cache for modern AI workloads.

\section{Motivation}
\label{sec:motivation}

\subsection{Limitations of Existing Works}
\label{sec:mot-workflow-adaptation}

\emph{Adaptivity} and \emph{transparency} are both desirable for an AI cache, yet they conflict, and existing works do not achieve both.

\textbf{\emph{Code-intrusive adaptations} compromise user transparency.}
Given the distinct policy preferences of different access patterns (Table~\ref{tab:access-patterns}), some existing works adapt through customized caching interfaces.
For example, Quiver exploits sample substitutability in shuffled training~\cite{quiver}, and NoPFS predicts training accesses from the shuffle seed~\cite{nopfs}.
Extending them to other phases requires new knowledge of how those phases access data and which cache behavior suits them, which conflicts with using the cache without workload-specific modifications or manual tuning.

\begin{figure}[t]
\centering
\includegraphics[width=\columnwidth]{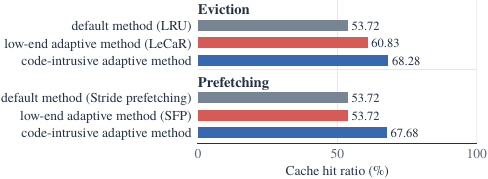}
\caption{Caching performance (cache hit ratio) of low-end and code-intrusive adaptations on ResNet-50 workflow
(W1 in Table~\ref{tab:eval-workflows}).
For code-intrusive, we manually select the most suitable prefetching or
eviction policy for each workflow phase according to its access pattern.}
\label{fig:mot-pattern-gap}
\end{figure}

\textbf{\emph{Low-end adaptations} compromise caching efficacy.}
Given the limitations of interface-based adaptations, other methods adopt simple \emph{heuristics} or lightweight learning that automatically adjust cache policies according to the observed access history.
For eviction, ARC balances recency and frequency~\cite{arc}, and LeCaR uses regret minimization to choose between LRU and LFU~\cite{lecar}; for prefetching, SFP analyzes access correlations between files with a Markov chain and adapts its prefetching window~\cite{sfp}.
Without workload hints, however, these methods handle microscopic variations within one pattern (e.g., \emph{skewed}), not macroscopic transitions across the patterns in Table~\ref{tab:access-patterns}.
As a result, their cache hit ratio is substantially below that of code-intrusive adaptations (Figure~\ref{fig:mot-pattern-gap}).

\textbf{Lessons learned: the need for LLM intelligence to attain high-end adaptivity.}
Given the limitations of both code-intrusive and low-end adaptations, what we need is a \emph{sufficiently flexible} caching method that works smoothly for arbitrary AI workloads.
Under the transparency requirement, the cache itself must infer the upper-level computation logic, and then the desired policies in Table~\ref{tab:access-patterns}, from low-level access sequences.
Given heterogeneous data semantics, intra-workflow dynamics, and the difficulty of telling random from skewed patterns, deterministic heuristics can hardly do so.
\emph{Large language models} (LLMs)~\cite{gpt3} have the potential to fill this gap: with their zero-shot problem-solving capability, they have been exploited in \emph{networking}~\cite{netllm} and \emph{databases}~\cite{dbot}.
We therefore explore how to exploit LLM intelligence in cache management for AI workloads.

\subsection{Challenges in Enforcing LLM-Based Adaptations}
\label{sec:mot-llm-opportunity}
\label{sec:mot-e2e-control}

\begin{figure}[t]
\centering
\includegraphics[width=\columnwidth]{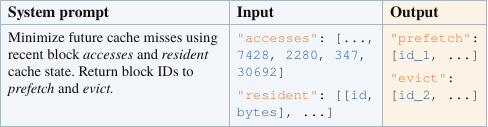}
\caption{Input/output interface of our end-to-end cache-control prototype.}
\label{fig:mot-prototype-io}
\end{figure}

While promising, exploiting LLM intelligence in adaptive cache management is non-trivial.
Here we empirically elaborate the challenges therein.
As shown in Figure~\ref{fig:mot-prototype-io}, we build a straightforward prototype atop JuiceFS that exploits LLM intelligence in an \emph{end-to-end} manner: given a JSON input with recent block accesses and the resident cache state, the model outputs the block IDs to prefetch or evict after every read request.
We measure\footnote{For decision quality, we use a reduced-cost setup to avoid the prohibitive reasoning delay: six models with thinking disabled, a 256\,MiB cache, a 200-access history, and a simpler PDEBench workflow (Section~\ref{sec:eval-e2e}). Quality runs share a warm start, exclude inference delay, and test prefetching and eviction separately.} its behavior with Qwen3.8 Flash on a small-scale diffusion LoRA development workflow (Section~\ref{sec:eval-e2e}), a 1,024-access history, and a 1\,GiB cache.
The results reveal three challenges.

\begin{figure}[t]
\centering
\includegraphics[width=\columnwidth]{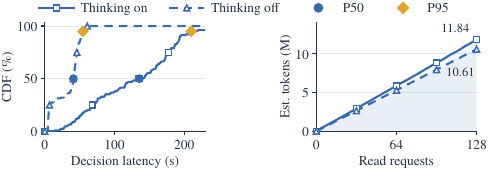}
\caption{Qwen3.8 Flash overhead over 128 calls per configuration.
Left: latency CDF. Right: token cost including input and output tokens.
Both include the two cases of thinking-on and thinking-off.} %
\label{fig:mot-naive-overhead}
\end{figure}

\textbf{Challenge-I: long reasoning delay.}
Inference is far too slow for real-time control on the read path.
Cloud-storage clients can issue hundreds of read requests per second: \emph{Bodner et al.} report 250--350 asynchronous read requests per second from a single client instance accessing S3~\cite{bodner2025serverless}.
Synchronous cache control makes read requests wait for model decisions, so inference must keep pace with frequent data accesses.
Yet Qwen3.8 Flash takes 136\,s at P50 and 210\,s at P95 with thinking on, and 41\,s at P50 with thinking off and capped output (Figure~\ref{fig:mot-naive-overhead}, left).

\textbf{Challenge-II: high token cost.}
The prototype repeatedly submits the updated history and resident state (Figure~\ref{fig:mot-prototype-io}).
Over 128 read requests, total tokens reach 11.84\,M with thinking on and 10.61\,M with thinking off, so the full diffusion LoRA workflow (about 60k read requests) would require about 5 billion tokens.

\begin{figure}[t]
\centering
\includegraphics[width=\columnwidth]{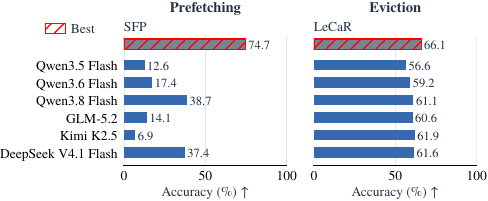}
\caption{Cache-action quality.
Prefetching accuracy is the fraction of prefetched blocks used before eviction or the end of the workload.
Eviction accuracy is the fraction of evicted blocks next requested beyond the
Belady boundary~\cite{lrb} or not requested again in the remaining trace.}
\label{fig:mot-naive-quality}
\end{figure}

\textbf{Challenge-III: poor decision quality.}
Despite their reasoning capability, LLMs make less accurate cache decisions than existing heuristics (Figure~\ref{fig:mot-naive-quality}).
The prefetching accuracy of the best LLM is 48.2\% lower than that of SFP, and even the highest model CHRs stay below the baselines: 80.8\% (DeepSeek) versus 82.3\% (SFP) for prefetching, and 82.6\% (Kimi) versus 83.9\% (LeCaR) for eviction.
A key reason is the large decision space: the LLM must choose prefetching targets from 1,963 candidates and eviction targets from 63.
Potential LLM hallucination further lowers decision quality.

In summary, to bring LLM-powered adaptive caching into practice, we need to shorten the reasoning delay, reduce the token cost, and also improve the decision quality.

\section{Solution}
\label{sec:design}

\subsection{Overview}
\label{sec:architecture}

\begin{figure}[!t]
\centering
\includegraphics[width=\columnwidth]{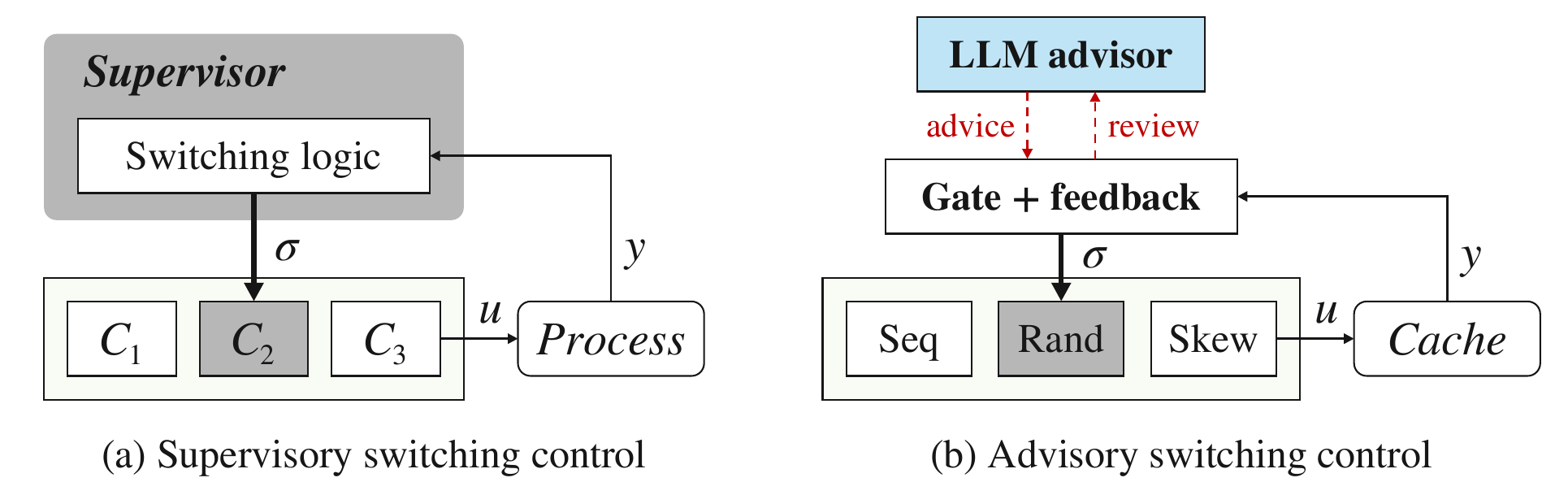}
\caption{Supervisory versus advisory switching control.
(a)~A trusted supervisor switches ($\sigma$) among controllers $C_i$ based on output $y$.
(b)~In \chiron{}, the LLM only advises; a deterministic harness gates its advice, switches among policy tools, and returns feedback for review.
Thick arrows carry switching authority; dashed arrows are asynchronous.}
\label{fig:principle}
\end{figure}

\noindent\textbf{Design insight: \emph{advisory switching control}.}
We propose \chiron{}, a fast, cheap, and effective cache system for AI workloads.
Instead of end-to-end LLM control, \chiron{} borrows \emph{supervisory switching control} from control theory~\cite{morse1996supervisory,liberzon2003switching}: a supervisor recognizes the operating regime and switches among pre-designed controllers, each encapsulating expert knowledge for one regime (Figure~\ref{fig:principle}a).
Replacing continuous tuning with discrete switching responds quickly to large regime changes and keeps the system modular and controllable, balancing LLM flexibility with reliability and efficiency.
Unlike a trusted supervisor, an LLM may hallucinate.
We thus adopt \emph{advisory switching control} (Figure~\ref{fig:principle}b): the LLM only recommends a policy, while a deterministic harness decides whether to adopt it and returns cache feedback so the LLM can revise its advice.

\begin{figure}[!t]
\centering
\includegraphics[width=\columnwidth]{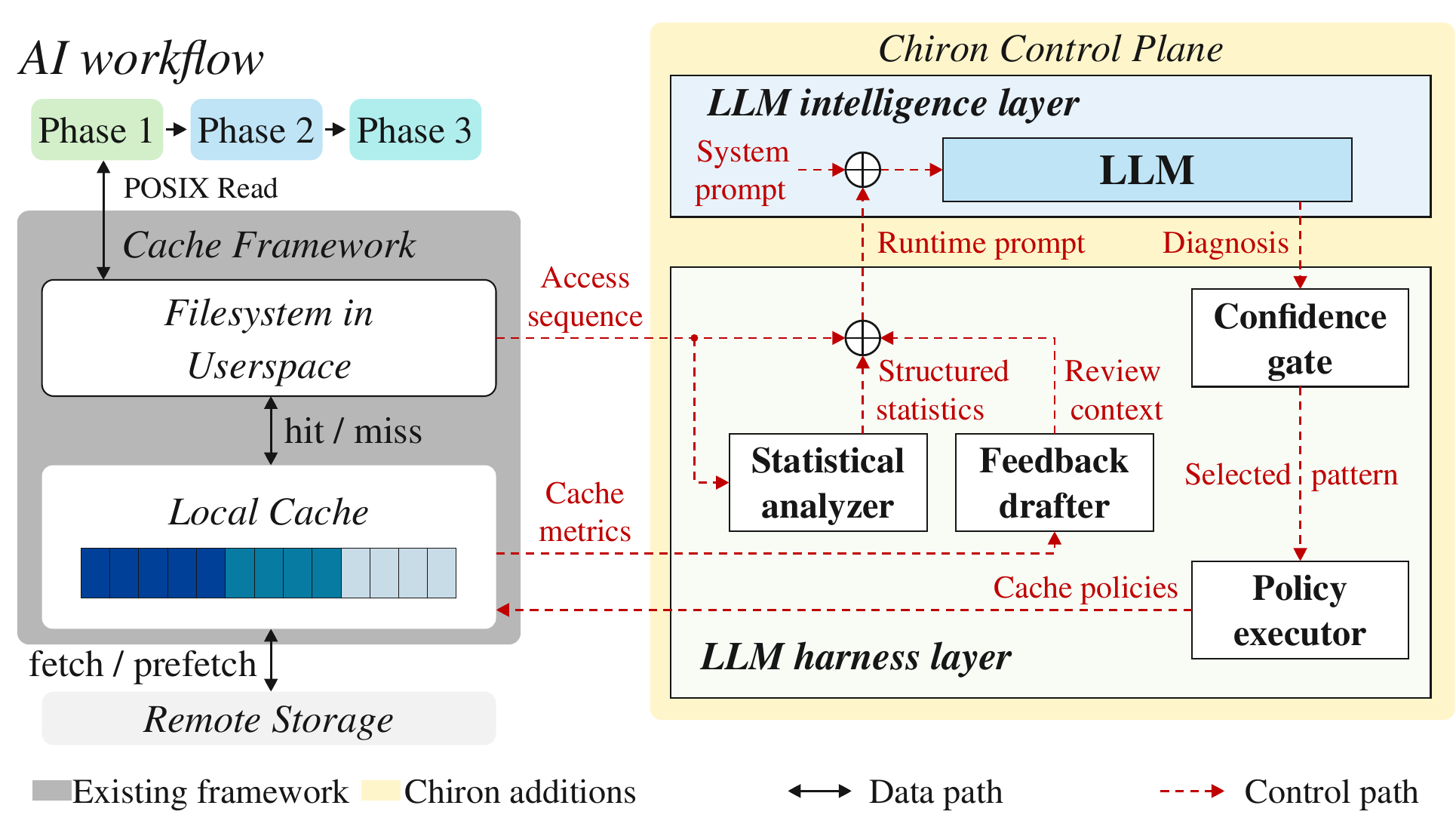}
\caption{Chiron architecture.} %
\label{fig:overview}
\end{figure}

\noindent\textbf{Architecture and workflow.}
As Figure~\ref{fig:overview} shows, \chiron{} has two layers.
In the \emph{LLM intelligence layer}, an \emph{executor-switching advisor} asynchronously diagnoses the access pattern: a \emph{runtime prompt} packages recent FUSE accesses, precomputed statistics, and review context, and the advisor combines it with a fixed system prompt and returns a short JSON with a pattern and its confidence.
In the \emph{LLM harness layer}, the \emph{statistical analyzer} prepares the statistics; the \emph{policy executor} orchestrates one policy tool per pattern and adopts a recommendation only if it passes a \emph{confidence gate}, after which the active executor makes every prefetching, admission, and eviction decision locally; and the \emph{feedback drafter} monitors cache outcomes and requests routine or early checks with a drafted context.

\noindent\textbf{Why fast, cheap, and effective.}
\chiron{} is \emph{fast} because inference stays off the read path and outputs are short; this asynchrony suffices because access patterns persist across many reads.
It is \emph{cheap} because the LLM selects policies instead of listing blocks and is invoked only when routine checks or feedback call for review.
It is \emph{effective} because choosing among expert-designed policies shrinks the decision space from thousands of candidate blocks to three patterns, while deterministic execution, gated adoption, and feedback-triggered review contain hallucination.

\subsection{LLM Intelligence Layer}
\label{sec:brain}
\label{sec:diagnosis}

A key part of the \chiron{} design is the input and output format of the LLM advisor, summarized in Figure~\ref{fig:brain-design}.

\subsubsection{Succinct yet Informative LLM Input}
\label{sec:brain-input}
\label{sec:input-analyzer}
\begin{figure*}[!t]
\centering
\includegraphics[width=\textwidth]{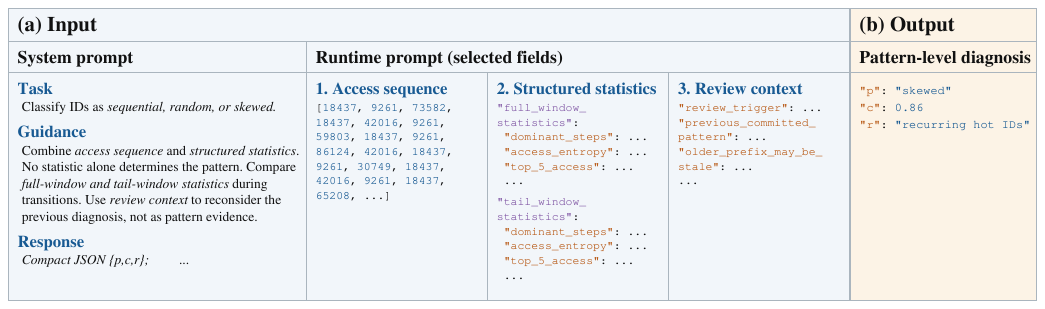}
\caption{The diagnosis interface combines fixed instructions and three groups of runtime input with a compact pattern-level response.
Only key instructions and selected \emph{runtime prompt} fields are shown; ellipses omit details given in Appendix~A.
The response is illustrative.}
\label{fig:brain-design}
\label{fig:diagnosis-json}
\end{figure*}

The input must support accurate judgment at low token cost.
We therefore combine expert knowledge in a \emph{system prompt} with runtime cache status in a \emph{runtime prompt}.
The system prompt describes the diagnosis task, which encodes the pattern knowledge of Table~\ref{tab:access-patterns}, explains how to interpret the runtime prompt, and constrains the output format (Section~\ref{sec:brain-output}).
The runtime prompt has three parts, numbered 1--3 in Figure~\ref{fig:brain-design}; Appendix~A gives both prompts in full.

\begin{itemize}
  \item \emph{Access sequence.}
Diagnosis depends on access order and recurrence.
\texttt{Access sequence} preserves the observed sequence so the advisor can inspect traversal and repeated objects directly.
The harness copies a recent window\footnote{The window size is fixed to 1,000: a modern AI workload (Table~\ref{tab:eval-workflows}) usually issues over 1,000 accesses in each phase, and its programming logic often persists during such a window.} from each cache unit, a file or directory managed together, keeping submitted evidence fixed while reads continue.

  \item \emph{Structured statistics.}
Deriving stride and reuse statistics from raw IDs is inefficient and can produce unreliable judgments.
\texttt{Structured statistics} therefore supplies precomputed values from the statistical analyzer, such as \texttt{access\_entropy}, which provide cues for sequential, random, and skewed patterns at two window scales (Section~\ref{sec:statistical-analyzer}).

  \item \emph{Review context.}
An accepted diagnosis may be incorrect or become outdated as accesses change.
The feedback drafter therefore requests a new diagnosis when cache feedback suggests a mismatch (Section~\ref{sec:reflection}).
To interpret this request, the advisor needs the earlier decision and the reason for reconsidering it, which \texttt{Review context} supplies.
\end{itemize}

\subsubsection{Token-efficient LLM Output}
\label{sec:brain-output}
Listing actions for individual objects would lengthen the response as the working set grows.
We instead return the compact pattern-level diagnosis in Figure~\ref{fig:brain-design}, which limits generation work and lets the policy executor reuse an accepted diagnosis and choose objects locally.
The pattern (\texttt{p}) specifies which prefetching and eviction strategies to consider for adoption (Section~\ref{sec:policy-executor}).
Since the advisor may hallucinate a pattern unsupported by the evidence, the \emph{system prompt} also asks for a self-assessed confidence score (\texttt{c}) in $[0,1]$, which the confidence gate uses (Section~\ref{sec:confidence-gate}).
An optional brief reason (\texttt{r}) can be enabled during debugging.

\subsection{LLM Harness Layer}
\label{sec:harness}

The LLM harness layer prepares statistical references, executes timely actions, defers unsupported policy changes, and reconsiders diagnoses as accesses change, through the four components below.

\subsubsection{Statistical Analyzer for Reliable Evidence}
\label{sec:statistical-analyzer}
The statistical analyzer, essentially a tool for LLM use, computes the \texttt{Structured statistics} introduced in Section~\ref{sec:brain-input} from the \texttt{Access sequence} and supplies them to the \emph{runtime prompt}.
These statistics serve as references for assessing how closely the sequence matches the three access patterns.
Figure~\ref{fig:brain-design} shows three example metrics, which provide cues for sequential, random, and skewed access, respectively.
For a sequence $X=(x_1,\ldots,x_n)$, \texttt{dominant\_steps} counts each adjacent ID difference $\delta=x_{i+1}-x_i$ and reports up to five most frequent differences with their counts, exposing repeated traversal steps.
To compute \texttt{access\_entropy}, the analyzer counts occurrences $c_u$ of each distinct ID $u$ and normalizes them as $p_u=c_u/n$.
With $m$ distinct IDs, the metric is $E=-\frac{\sum_u p_u\log_2 p_u}{\log_2 m}$, with $E=0$ for $m\leq1$; larger values indicate more evenly distributed access frequencies.
Finally, \texttt{top\_5\_access} sums the counts of the five most frequent IDs and divides by $n$: $S_5=\sum_{j=1}^{\min(5,m)}c_{[j]}/n$, where $c_{[j]}$ is the $j$th largest count.
Appendix~A.2 gives the complete set of metrics.
The analyzer also reports these statistics separately for the full window and its recent tail.
This helps the advisor identify recent behavior during pattern transitions, when accesses from the previous phase may dilute the evidence in the full-window summary.

\subsubsection{Deterministic Policy Executor for Fast Action}
\label{sec:policy-executor}
Cache contents can change during inference, making object choices from a diagnosis snapshot outdated.
The deterministic policy executor therefore applies the active policy to current accesses and cache state, selecting prefetching and eviction actions locally.
The sequential executor uses stride prefetching; the random executor disables pattern-based prefetching and rejects new admissions when full to retain the cached subset; the skewed executor combines neighborhood prefetching with LRU eviction.
An accepted pattern change switches the active executor.
Reads initially use the framework's default policy, and pending diagnoses leave the active policy unchanged.

\subsubsection{Confidence Gate for Hallucination Avoidance}
\label{sec:confidence-gate}
\begin{figure}[!t]
\centering
\includegraphics[width=\columnwidth]{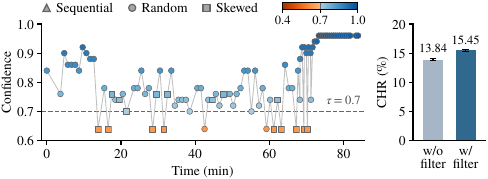}
\caption{Patterns and confidence scores of diagnoses from Qwen3-Max (released in September 2025 and relatively dated compared with models we evaluate in Section~\ref{sec:evaluation}) (left) and CHR with and without a confidence filter (right) during the training phase of the ResNet-50 workflow.}
\label{fig:confidence-gate-motivation}
\end{figure}

To control inference cost, some deployments may use models with limited reasoning capability as advisors.
Such advisors may hallucinate an access pattern, leading the executor to adopt an unsuitable cache policy.
We observe that hallucinated diagnoses often coincide with low model confidence, so the score \texttt{c} (Section~\ref{sec:brain-output}) can identify some of them before adoption.
For example, in the training phase of the ResNet-50 workflow with randomly shuffled accesses (Section~\ref{sec:eval-e2e}), filtering diagnoses with $c<0.7$ while retaining the existing policy improves CHR from $13.84\%$ to $15.45\%$ (Figure~\ref{fig:confidence-gate-motivation}).
The confidence gate therefore requires \(c\geq\tau\) before adopting a proposed pattern \(p\), and otherwise retains the existing policy.
Since a simple filter does not correct a wrong diagnosis, the harness collects more accesses and retries sooner than the routine interval, allowing fresh evidence to help correct the diagnosis.

\subsubsection{Feedback Reflection for Fewer LLM Calls}
\label{sec:reflection}
\begin{figure}[!t]
\centering
\includegraphics[width=\columnwidth]{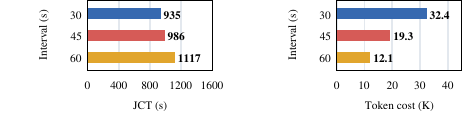}
\caption{Diagnosis schedule tradeoff in the ResNet-50 workflow (W1 in Table~\ref{tab:eval-workflows}): JCT and token cost under different diagnosis intervals.}
\label{fig:reflection-motivation}
\end{figure}

Invoking the LLM periodically creates a tradeoff: frequent calls consume tokens even when the last accepted diagnosis still matches the access pattern, while longer intervals can delay adaptation when the pattern changes.
Figure~\ref{fig:reflection-motivation} illustrates this tradeoff in a replay of the ResNet-50 workflow, using a quarter of the dataset and cache budget.
The feedback drafter therefore monitors runtime I/O with lightweight local rules and requests an early LLM review when necessary, allowing less frequent routine reviews.

The drafter uses CHR fluctuations as a sign that the current pattern may no longer fit.
For example, a random-to-skewed transition can raise CHR through increased reuse, yet the random executor's fixed cached subset can exclude new hot objects that the skewed executor would admit.
Let \(h_t\) be CHR in round \(t\), and \(H_t=\alpha h_t+(1-\alpha)H_{t-1}\) its exponentially weighted moving average (EWMA), with smoothing weight \(\alpha\).
A review is requested when \(|h_t - H_{t-1}| \geq \theta\), where \(\theta\) is the fluctuation threshold.
The drafter then sends the CHR fluctuation and the previous diagnosis to the advisor in the \texttt{Review context} (Section~\ref{sec:brain-input}), alerting it that the earlier diagnosis may be outdated.

\section{Implementation}
\label{sec:implementation}

We implement \chiron{} as an extension to JuiceFS, comprising a client-side controller written in Go and an LLM diagnosis server written in Python, with about 4,000 LoC of Go and 400 LoC of Python.
The controller implements the policy executor and the feedback drafter of the LLM harness layer; the server implements the statistical analyzer and hosts the executor-switching advisor.

\noindent\textbf{Diagnosis service and asynchronous interaction.}
The server connects the advisor to the LLM through LangChain's chat-model interface and exposes it to the controller via a FastAPI endpoint.
To keep cache control asynchronous with LLM inference, the controller snapshots the access history and review context, then launches a background goroutine for the HTTP exchange while foreground reads continue under the active policy.
The server constructs the runtime prompt from this input and from statistics computed on the access history, and returns the diagnosis as JSON.

\noindent\textbf{Client-side controller.}
To obtain the access sequence in Figure~\ref{fig:overview}, we add a hook to JuiceFS's FUSE \texttt{Read()} handler that collects each demand request's inode, offset, and length.
The controller converts this information into block IDs using the file-to-block mapping in metadata and records them in access order.
The feedback drafter uses a background goroutine to periodically compute CHR fluctuations and update the review context.
The policy executor calls \texttt{VFS.Read()} to fetch prefetch targets and removes selected cache files through the chunk store, which calls \texttt{os.Remove()}.
Prefetch reads bypass the foreground hook to keep speculative accesses out of the access history.

\noindent\textbf{Scalability.}
To maintain effective cache management and controlled diagnosis cost as datasets, caches, and access histories grow, \chiron{} keeps LLM input and output lengths bounded.
The advisor uses a recent window of access history and returns a compact pattern-level diagnosis, so asynchronous inference can overlap with foreground reads as scale increases.
Bounded inputs and outputs also control the token cost of each diagnosis, and the feedback drafter reduces the frequency of reviews.

\section{Evaluation}
\label{sec:evaluation}

We evaluate Chiron from five perspectives.
First, we measure its end-to-end effectiveness across multi-phase and single-phase workloads (\S\ref{sec:eval-e2e}).
Second, we assess the advantage of its pattern-aware eviction and prefetching over fixed policies (\S\ref{sec:eval-mechanism}).
Third, we test the necessity of structured statistics, confidence gating, and feedback reflection (\S\ref{sec:eval-ablation}).
Fourth, we analyze how LLM capability affects Chiron (\S\ref{sec:eval-models}).
Finally, we assess its cache performance at large dataset scale in simulation (\S\ref{sec:eval-data-scaling}).

\subsection{End-to-End Performance}
\label{sec:eval-e2e}

\noindent\textbf{Workload and hardware setup.}
We use nine representative AI workloads (Table~\ref{tab:eval-workflows}): three \emph{multi-phase workflows}, which involve both pattern heterogeneity and dynamicity, and six \emph{single-phase tasks}, which involve only pattern heterogeneity.
Computations run on a server with eight A800 GPUs and an SSD-based cache, and datasets are stored remotely on S3 at a measured network latency of roughly 250~ms.
To avoid trivial cases where the entire dataset can be cached without prefetching or eviction, we set a modest cache size for each experiment based on the original dataset size. The dataset and cache sizes\footnote{For trial efficiency we adopt relatively small dataset and cache sizes; such scale-down methods are commonly used in the literature~\cite{zhao2023silod,khan2023shade}, and Section~\ref{sec:eval-data-scaling} evaluates larger scale in simulation.} are listed in Table~\ref{tab:eval-workflows}.

\begin{table*}[!t]
\centering
\scriptsize
\setlength{\tabcolsep}{1.5pt}
\renewcommand{\arraystretch}{1.12}
\begin{tabularx}{\textwidth}{|>{\centering\arraybackslash}p{0.055\textwidth}|
                >{\centering\arraybackslash}X|
                >{\centering\arraybackslash}p{0.205\textwidth}|
                >{\centering\arraybackslash}p{0.205\textwidth}|
                >{\centering\arraybackslash}p{0.19\textwidth}|}
\hline
\textbf{ID} & \textbf{Workloads} & \textbf{Dataset} & \textbf{Dataset Size (Cache Budget)} & \textbf{Access Pattern} \\
\hline
W1 & ResNet-50 workflow~\cite{resnet} & ImageNet subset~\cite{hf-imagenet} & 1.395~GiB (1~GiB) & Sequential, random, skewed \\
\hline
W2 & Minimal-FNO workflow~\cite{pdebench} & PDEBench~\cite{hf-pdebench} & 7.67~GiB (4~GiB) & Sequential, random, skewed \\
\hline
W3 & Stable Diffusion 1.5 workflow~\cite{hf-sd15} & Mercity LAION subset~\cite{hf-mercity} & 3.87~GiB (1~GiB) & Random, sequential \\
\hline
T1 & Qwen tokenization task~\cite{hf-qwen25} & C4 shards~\cite{hf-c4} & 2.37~GiB (2~GiB) & Sequential \\
\hline
T2 & Qwen2.5-3B weight loading task~\cite{hf-qwen25} & Qwen2.5-3B weights~\cite{hf-qwen25} & 5.75~GiB (4~GiB) & Sequential \\
\hline
T3 & ViT-Base fine-tuning task~\cite{dosovitskiy2021vit} & COCO 2014 subset~\cite{hf-coco2014} & 6.09~GiB (1~GiB) & Random \\
\hline
T4 & Wav2Vec2 fine-tuning task~\cite{wav2vec2} & Speech Commands v0.02~\cite{hf-speech} & 2.48~GiB (1~GiB) & Random \\
\hline
T5 & DLRM training task~\cite{dlrm} & Criteo\_x1 trace~\cite{hf-criteo} & 1.99~GiB (1~GiB) & Skewed \\
\hline
T6 & RAG retrieval task~\cite{rag} & Wikipedia / MMLU~\cite{hf-wikipedia,hf-mmlu} & 3.04~GiB (0.5~GiB) & Skewed \\
\hline
\end{tabularx}
\caption{Workloads named by model in the end-to-end experiment. IDs label the
workloads in Figures~\ref{fig:eval-e2e-phase} and~\ref{fig:eval-e2e-stationary}.}
\label{tab:eval-workflows}
\end{table*}

\begin{figure}[!t]
\centering
\includegraphics[width=\columnwidth]{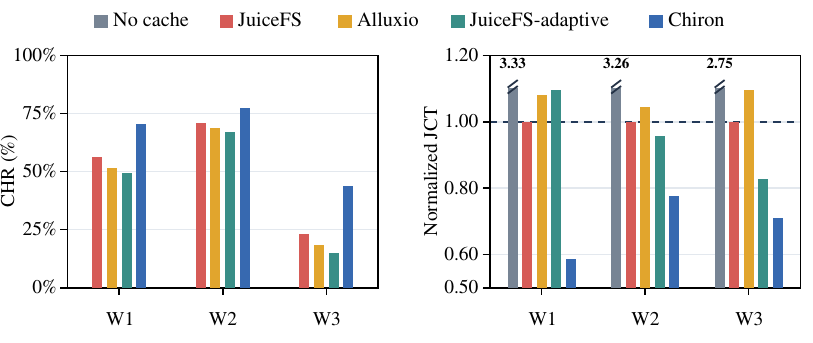}
\caption{Five-policy comparison on multi-phase workflows W1--W3 in
Table~\ref{tab:eval-workflows}.
No-cache has zero CHR in all three workflows and is omitted from the CHR plot.
JCT is normalized to JuiceFS within each workflow.
}
\label{fig:eval-e2e-phase}
\end{figure}

\noindent\textbf{Baselines.}
We compare Chiron with four baselines: (1)~No-cache, which fetches demand data from remote storage without retaining read blocks; (2)~JuiceFS, which uses \emph{sequential} block prefetching and \emph{2-random} eviction, repeatedly evicting the older of two sampled blocks~\cite{juicefs-2random}; (3)~Alluxio\footnote{For a fair comparison, we implement the cache-management logic of Alluxio (written in Java) over JuiceFS (written in Go).}, using its default LRU eviction without prefetching; and (4)~JuiceFS-Adaptive, a variant of JuiceFS using LeCaR for eviction and SFP for prefetching.
Later micro-benchmarks use other prefetching and eviction baselines.

\noindent\textbf{Chiron setup.}
By default, the LLM is DeepSeek-V4-flash; Section~\ref{sec:eval-models} tests other models.
We set the confidence threshold $\tau$ to 0.70 (\S\ref{sec:confidence-gate}), and for feedback reflection (\S\ref{sec:reflection}) the EWMA smoothing weight $\alpha$ to 0.35 and the absolute CHR fluctuation threshold $\theta$ to 0.18.
Appendix~B analyzes the sensitivity to this threshold.

\noindent\textbf{Metrics.}
We report block-level cache hit ratio (CHR) and end-to-end job completion time (JCT), which are common in the literature~\cite{khan2023shade,quiver}.
Micro-benchmarks also report token cost and pattern diagnosis accuracy.

\begin{figure}[!t]
\centering
\includegraphics[width=\columnwidth]{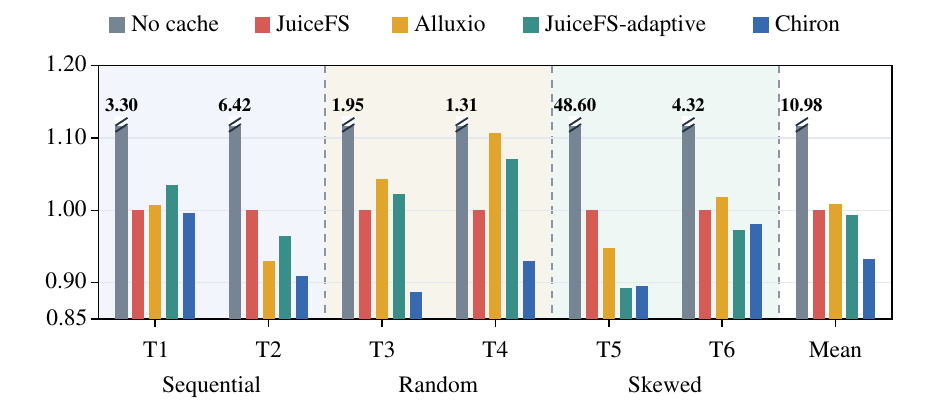}
\caption{JCT on single-phase tasks T1--T6 in Table~\ref{tab:eval-workflows},
normalized to JuiceFS for each task.}
\label{fig:eval-e2e-stationary}
\end{figure}

\noindent\textbf{Overall improvements.}
As Figure~\ref{fig:eval-e2e-phase} shows, adopting a cache substantially reduces JCT.
More importantly, Chiron has the highest CHR and lowest JCT on multi-phase workflows, reducing JCT by 41.2\%, 18.8\%, and 14.3\% over the best baseline on the three workflows.
This confirms the power of LLM intelligence in adjusting the caching policy at runtime.
On single-phase tasks (Figure~\ref{fig:eval-e2e-stationary}), Chiron reduces JCT by an average of 5.22\% over the best baseline on four of six tasks, and is near parity on the other two, skewed workloads that JuiceFS-Adaptive already suits.
Chiron thus yields generally effective caching across AI workloads.

\subsection{Microscopic Performance Deep Dive}
\label{sec:eval-mechanism}

We next examine prefetching and eviction separately on the Stable Diffusion 1.5 workflow (W3 in Table~\ref{tab:eval-workflows}), with the setup of Section~\ref{sec:eval-e2e} unless otherwise specified.

\begin{figure}[!t]
\centering
\subcaptionbox{Eviction\label{fig:eval-mechanisms-eviction}}{%
\includegraphics[width=\columnwidth]{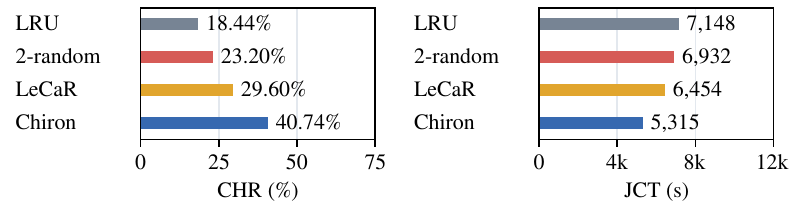}}
\par\vspace{0.5\baselineskip}
\subcaptionbox{Prefetch\label{fig:eval-mechanisms-prefetch}}{%
\includegraphics[width=\columnwidth]{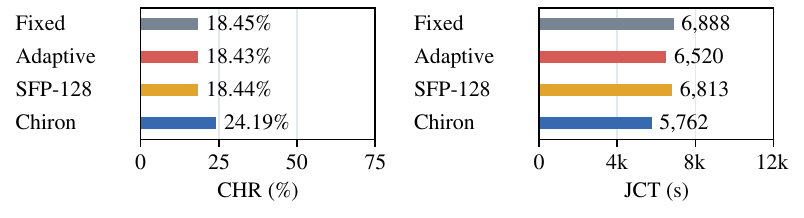}}
\caption{Eviction and prefetch comparisons on the Stable Diffusion 1.5 workflow (W3 in Table~\ref{tab:eval-workflows}).}
\label{fig:eval-mechanisms}
\end{figure}

\noindent\textbf{Eviction performance.}
We compare Chiron with LRU, LeCaR, and JuiceFS 2-random, with prefetching disabled and the same cache size in all arms.
Chiron has the highest CHR and reduces JCT by 17.65\% relative to the best baseline (Figure~\ref{fig:eval-mechanisms-eviction}), which confirms that its eviction policy is better than static or low-end adaptive methods.

\noindent\textbf{Prefetching performance.}
We compare Chiron with sequential prefetching at fixed depth, sequential prefetching at adaptive depth~\cite{juicefs-cache}, and SFP~\cite{sfp}, with the same cache size and LRU eviction in all arms.
Chiron outperforms all three, improving CHR by 31.11\% (24.19\% versus 18.45\%) and JCT by 11.63\% (Figure~\ref{fig:eval-mechanisms-prefetch}).

\subsection{Ablation Study}
\label{sec:eval-ablation}

Section~\ref{sec:design} adopts multiple techniques to make Chiron fast, cheap, and effective, such as \emph{structured statistics}, \emph{confidence gating}, and \emph{feedback reflection}.
We evaluate each of them here, on diverse workflows for representativeness.

\begin{figure*}[!t]
\centering
\subcaptionbox{Structured statistics\label{fig:eval-raw}}{\includegraphics[height=.12\textwidth]{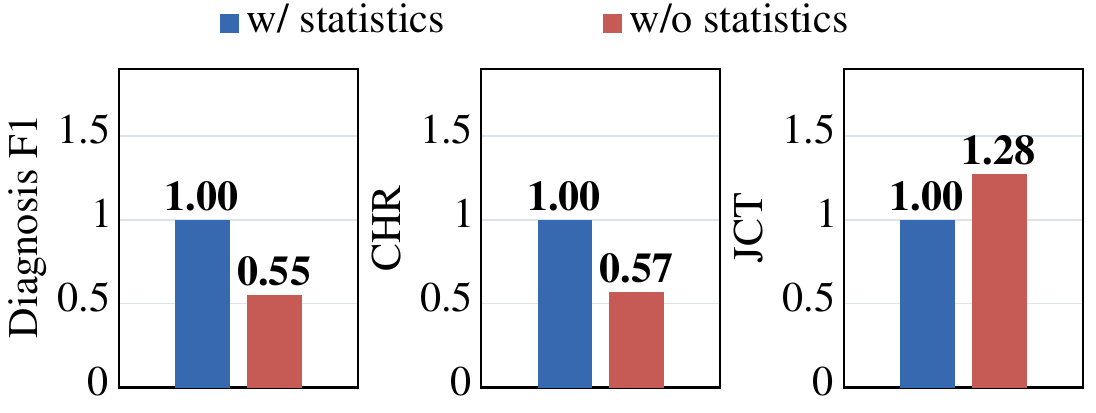}}\hfill
\subcaptionbox{Confidence gate\label{fig:eval-confidence-gate}}{\includegraphics[height=.12\textwidth]{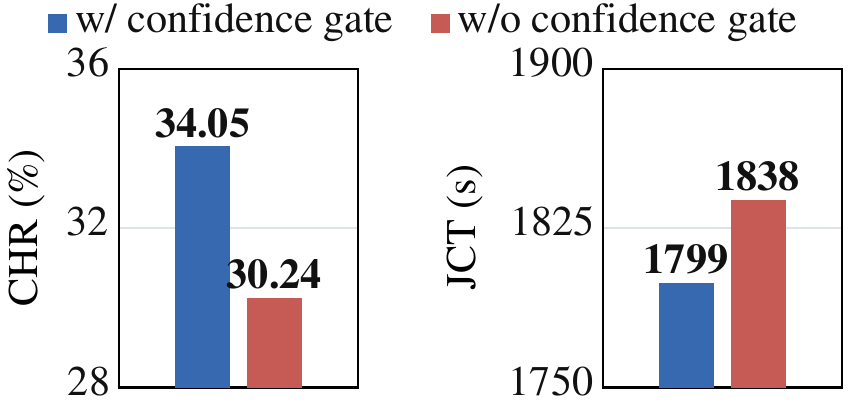}}\hfill
\subcaptionbox{Feedback reflection\label{fig:eval-reflection}}{\includegraphics[height=.12\textwidth]{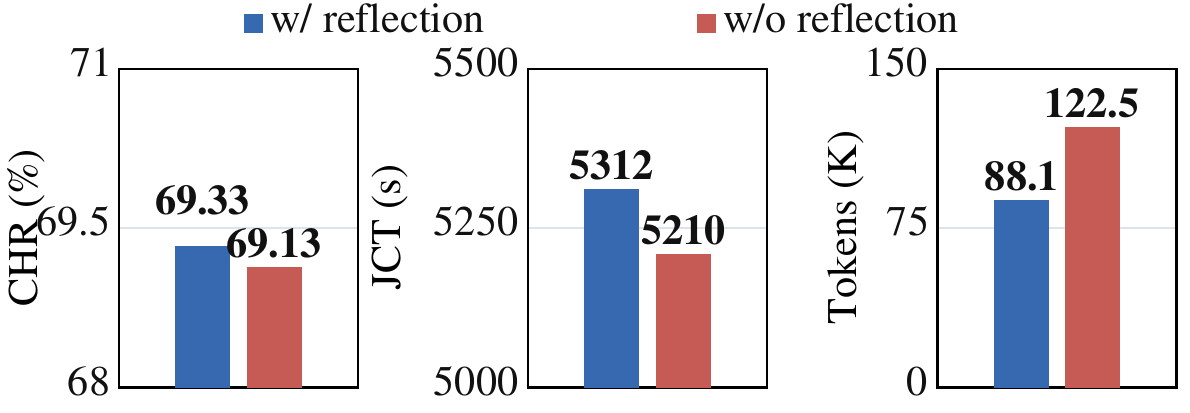}}
\caption{Ablation studies. Structured statistics are evaluated on the Stable Diffusion 1.5 workflow (W3 in Table~\ref{tab:eval-workflows}), normalized to full Chiron; confidence gating and feedback reflection are evaluated on the ResNet-50 workflow trace.}
\label{fig:eval-ablation}
\end{figure*}

\noindent\textbf{Necessity of structured statistics.}
We compare Chiron with a variant that has no structured statistics in its prompt, on the Stable Diffusion 1.5 workflow.
Without structured statistics, the F1 score of pattern diagnosis degrades by 45\% and CHR falls by 43\% (Figure~\ref{fig:eval-raw}), so these statistics improve both pattern recognition and end-to-end performance.

\noindent\textbf{Necessity of confidence gate.}
On the ResNet-50 workflow\footnote{We used a faster-switching variant here which has approximately one-tenth as many reads per phase and twice as many pattern switches as the original workflow.}, enabling the gate raises CHR by 12.60\% in relative terms and reduces JCT from 1,838.5 to 1,799.4~s (Figure~\ref{fig:eval-confidence-gate}).
Confidence gating is thus necessary to handle unstable LLM decisions.

\noindent\textbf{Necessity of feedback reflection.}
Also on the ResNet-50 workflow, we compare LLM calls once per 30~s \emph{without reflection} and once per 45~s \emph{with reflection}, which allows early calls triggered by feedback.
With reflection, Chiron attains comparable CHR and JCT with 28.1\% fewer tokens (Figure~\ref{fig:eval-reflection}).

\subsection{Sensitivity Analysis to LLM Capability}
\label{sec:eval-models}

\begin{figure}[!t]
\centering
\includegraphics[width=\columnwidth]{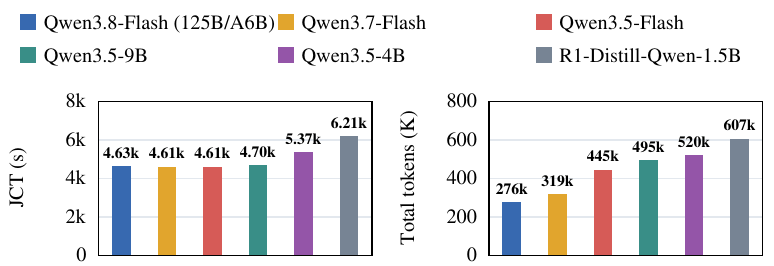}
\caption{Diagnosis-model sensitivity on the ResNet-50 workflow
replay.}
\label{fig:eval-models}
\end{figure}

We replay the ResNet-50 workflow with six models: Qwen3.8-Flash, Qwen3.7-Flash, Qwen3.5-Flash, Qwen3.5-9B, Qwen3.5-4B, and R1-Distill-Qwen-1.5B.
Chiron yields similar caching performance from Qwen3.8-Flash to Qwen3.5-9B, with JCT between 4.61k and 4.70k~s, yet degrades markedly at 4B and below, with JCT of 5.37k~s for Qwen3.5-4B and 6.21k~s for R1-Distill-Qwen-1.5B (Figure~\ref{fig:eval-models}).
A modest LLM, larger than 4B, is thus desirable.
Token cost also grows monotonically as the LLM shrinks, from 276k to 607k tokens, because weaker models need more thinking tokens.

\subsection{Large-Scale Simulation}
\label{sec:eval-data-scaling}

We build an event-driven cache simulator to assess Chiron at large scale, simulating the ResNet-50 workflow on an expanded 1~TB ImageNet dataset. This matches the TB-scale sizes of widely used AI datasets such as the 1.31~TB full-resolution ImageNet-21K archive~\cite{imagenet21k-size} and the approximately 10~TB LAION-400M WebDataset~\cite{laion400m-size}.
Each access costs 2~ms on a cache hit and 250~ms on a miss, and cache sizes are 10\%, 25\%, 50\%, and 85\% of the dataset.
In a fidelity test, the simulator differs from end-to-end measurements by 2.82 percentage points of CHR on average and 4.42 at most.

\begin{figure}[!t]
\centering
\includegraphics[width=\columnwidth]{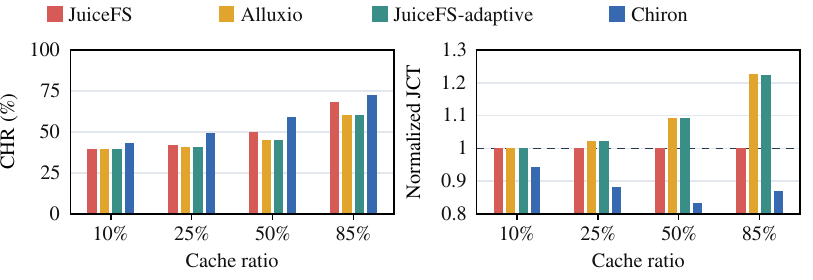}
\caption{Policy comparison in the large-scale dataset simulation. Cache
capacity is expressed as a fraction of dataset bytes. JCT is normalized to
JuiceFS at each capacity.}
\label{fig:eval-data-scaling}
\end{figure}

According to Figure~\ref{fig:eval-data-scaling}, Chiron has the highest CHR and lowest simulated JCT at every cache size, improving CHR by 3.61--8.79 percentage points and reducing JCT by 5.72--16.69\% over the best baseline.
The average LLM call delay stays below 1.3~s and is not on the critical path.
With bounded input and output lengths and asynchronous decisions, both independent of dataset size, Chiron can scale to production scenarios.

\section{Related Work}
\label{sec:related}

\noindent\textbf{Workload-aware caching for AI workloads.}
AI storage systems exploit training structure to reduce data-loading costs~\cite{cachew,hycache,sun2022solar,xu2022deep}.
Quiver~\cite{quiver} exploits sample substitutability, CoorDL~\cite{data-stalls-dnn} retains samples across shuffled epochs, and NoPFS~\cite{nopfs} predicts accesses from sampling seeds to coordinate caching and prefetching.
SHADE~\cite{khan2023shade} uses sample importance to guide sampling and caching, and SiloD~\cite{zhao2023silod} co-designs caching and cluster scheduling.
\chiron{} instead diagnoses observed file accesses to select policies across workflow phases, without sampling seeds or sample-importance scores.

\noindent\textbf{Adaptive and learning-based cache management.}
Adaptive policies infer useful cache behavior from access history~\cite{lirs,clockpro,beckmann2018lhd,yang2023gl}.
ARC~\cite{arc} balances recency and frequency, LeCaR~\cite{lecar} and CACHEUS~\cite{cacheus} learn to combine replacement experts, LRB~\cite{lrb} approximates Belady-style eviction, and SFP~\cite{sfp} learns file-access patterns for prefetching.
Learning-Augmented Heuristics~\cite{xia2026learning} also select cache parameters asynchronously while deterministic heuristics process requests; \chiron{} shares this separation, but uses LLM-based pattern diagnosis to coordinate prefetching, admission, and eviction.

\noindent\textbf{LLM-assisted systems management.}
LLMs increasingly support system diagnosis, tuning, and runtime decisions~\cite{singh2024panda,llmcompiler,chen2024autoos,huang2024e2etune}.
D-Bot~\cite{dbot} uses LLMs for database diagnosis, GPTuner~\cite{gptuner} extracts documentation knowledge to guide tuning, $\lambda$-Tune~\cite{giannakouris2025lambda} generates and evaluates database configurations, NetLLM~\cite{netllm} adapts LLMs to networking tasks, and An Expert in Residence~\cite{expert-os-tuning} explores online OS tuning.
PF-LLM~\cite{pfllm} separates offline assembly analysis from runtime hardware prefetching through LLM-generated hints; \chiron{} extends this division to online file caching, using reusable diagnoses and cache feedback to track changing access patterns.

\section{Conclusion}
\label{sec:conclusion}

AI workloads need a cache that adapts to diverse and changing access patterns without user hints.
\chiron{} provides it through advisory switching control: an LLM asynchronously diagnoses the access pattern and recommends a policy, while a deterministic harness gates this advice, executes expert-designed policies, and feeds cache outcomes back for review.
Implemented atop JuiceFS, \chiron{} reduces JCT by 14.3--41.2\% over the best baseline on multi-phase workflows.
The same division of work may suit other systems whose decisions are too frequent for an LLM to make directly.

\bibliographystyle{mlsys2025}
\bibliography{main}

\onecolumn
\appendix
\section{Diagnosis Prompt and Input Fields}
\label{sec:diagnosis-prompt}
\raggedbottom
\definecolor{snapshotkey}{HTML}{DC7026}
\definecolor{snapshotvalue}{HTML}{185A93}
\definecolor{snapshotparent}{HTML}{8056A4}

This appendix provides the full \emph{system prompt}, an example \emph{runtime prompt}, and definitions of its fields to accompany Section~\ref{sec:brain-input} and Figure~\ref{fig:brain-design}.
The prompt and its statistics fields use the terminology of the main text and Figure~\ref{fig:brain-design}.

\begin{lstlisting}[basicstyle=\ttfamily\footnotesize,
  columns=fullflexible,keepspaces=true,breaklines=true,
  breakatwhitespace=false,showstringspaces=false,
  frame=single,framerule=0.3pt,rulecolor=\color{black!35},
  framesep=5pt,literate={±}{{$\pm$}}1]
Classify an ordered block-ID sequence as exactly one pattern: "sequential", "random", or "skewed".

Definitions:
- sequential: ordered traversal or a stable stride; it need not use only ±1 steps.
- random: broadly dispersed accesses without stable order or concentrated reuse.
- skewed: repeated accesses concentrated on a comparatively small subset of IDs.

Use access sequence and structured statistics jointly to assess the candidate
patterns; no single statistic is a mandatory classification rule. Compare
full-window and tail-window statistics to distinguish recent behavior from
older accesses during a phase transition. Use review context to identify the
reason for review, the previous diagnosis, and potentially outdated accesses.
Treat this context as a reason to reconsider the diagnosis, not as evidence
for a pattern, and do not assume the previous diagnosis remains correct.

Normalize confidence to a score between 0 and 1 inclusive, not a percentage.
The confidence score expresses a model judgment, not a calibrated probability.
r is optional debugging output.
Return only valid JSON:
{"p":"sequential|random|skewed","c":<number from 0 to 1>,"r":"optional debug reason"}
\end{lstlisting}

\subsection{Runtime prompt}

The runtime prompt below uses the illustrative sequence $W=(0,2,\ldots,798)$ and a routine review of a previously accepted random pattern.
Only the sequence is abbreviated; all statistics and context fields are shown.

\begin{lstlisting}[basicstyle=\ttfamily\footnotesize,
  columns=fullflexible,keepspaces=true,breaklines=true,
  breakatwhitespace=false,showstringspaces=false,
  frame=single,framerule=0.3pt,rulecolor=\color{black!35},
  framesep=5pt,aboveskip=5pt,belowskip=5pt,
  literate=
  {"window"}{{\textcolor{snapshotparent}{"window"}}}{8}
  {"length"}{{\textcolor{snapshotkey}{"length"}}}{8}
  {"unique_ratio"}{{\textcolor{snapshotkey}{"unique\_ratio"}}}{14}
  {"access_entropy"}{{\textcolor{snapshotkey}{"access\_entropy"}}}{16}
  {"gini"}{{\textcolor{snapshotkey}{"gini"}}}{6}
  {"top_5_access"}{{\textcolor{snapshotkey}{"top\_5\_access"}}}{14}
  {"reuse_ratio"}{{\textcolor{snapshotkey}{"reuse\_ratio"}}}{13}
  {"short_reuse_ratio"}{{\textcolor{snapshotkey}{"short\_reuse\_ratio"}}}{19}
  {"mean_reuse_distance"}{{\textcolor{snapshotkey}{"mean\_reuse\_distance"}}}{21}
  {"adjacent_steps"}{{\textcolor{snapshotkey}{"adjacent\_steps"}}}{16}
  {"zero_ratio"}{{\textcolor{snapshotkey}{"zero\_ratio"}}}{12}
  {"pm1_ratio"}{{\textcolor{snapshotkey}{"pm1\_ratio"}}}{11}
  {"small_jump_ratio"}{{\textcolor{snapshotkey}{"small\_jump\_ratio"}}}{18}
  {"large_jump_ratio"}{{\textcolor{snapshotkey}{"large\_jump\_ratio"}}}{18}
  {"dominant_steps"}{{\textcolor{snapshotkey}{"dominant\_steps"}}}{16}
  {"delta"}{{\textcolor{snapshotkey}{"delta"}}}{7}
  {"count"}{{\textcolor{snapshotkey}{"count"}}}{7}
  {"recent_tail"}{{\textcolor{snapshotparent}{"recent\_tail"}}}{13}
  {"review_trigger"}{{\textcolor{snapshotkey}{"review\_trigger"}}}{16}
  {"previous_control_state"}{{\textcolor{snapshotkey}{"previous\_control\_state"}}}{24}
  {"previous_committed_pattern"}{{\textcolor{snapshotkey}{"previous\_committed\_pattern"}}}{28}
  {"older_prefix_may_be_stale"}{{\textcolor{snapshotkey}{"older\_prefix\_may\_be\_stale"}}}{27}
  {"recent_tail_length"}{{\textcolor{snapshotkey}{"recent\_tail\_length"}}}{20}
  {"periodic_retry"}{{\textcolor{snapshotvalue}{"periodic\_retry"}}}{16}
  {"stable"}{{\textcolor{snapshotvalue}{"stable"}}}{8}
  {"random"}{{\textcolor{snapshotvalue}{"random"}}}{8}
  {false}{{\textcolor{snapshotvalue}{false}}}{5}
  {0}{{\textcolor{snapshotvalue}{0}}}{1}
  {1}{{\textcolor{snapshotvalue}{1}}}{1}
  {2}{{\textcolor{snapshotvalue}{2}}}{1}
  {3}{{\textcolor{snapshotvalue}{3}}}{1}
  {4}{{\textcolor{snapshotvalue}{4}}}{1}
  {5}{{\textcolor{snapshotvalue}{5}}}{1}
  {6}{{\textcolor{snapshotvalue}{6}}}{1}
  {7}{{\textcolor{snapshotvalue}{7}}}{1}
  {8}{{\textcolor{snapshotvalue}{8}}}{1}
  {9}{{\textcolor{snapshotvalue}{9}}}{1}]
Access sequence: [0,2,4,...,794,796,798]
Structured statistics:
{"window": {
  "length": 400, "unique_ratio": 1.0, "access_entropy": 1.0,
  "gini": 0.0, "top_5_access": 0.013, "reuse_ratio": 0.0,
  "short_reuse_ratio": 0.0, "mean_reuse_distance": 400.0,
  "adjacent_steps": {
    "zero_ratio": 0.0, "pm1_ratio": 0.0,
    "small_jump_ratio": 1.0, "large_jump_ratio": 0.0,
    "dominant_steps": [{"delta": 2, "count": 399}]}},
 "recent_tail": {
  "length": 200, "unique_ratio": 1.0, "access_entropy": 1.0,
  "gini": 0.0, "top_5_access": 0.025, "reuse_ratio": 0.0,
  "short_reuse_ratio": 0.0, "mean_reuse_distance": 200.0,
  "adjacent_steps": {
    "zero_ratio": 0.0, "pm1_ratio": 0.0,
    "small_jump_ratio": 1.0, "large_jump_ratio": 0.0,
    "dominant_steps": [{"delta": 2, "count": 199}]}}}
Review context:
{"review_trigger": "periodic_retry", "previous_control_state": "stable",
 "previous_committed_pattern": "random", "older_prefix_may_be_stale": false,
 "recent_tail_length": 200}
\end{lstlisting}

\subsection{Runtime prompt field definitions}
\label{sec:input-field-definitions}

The \emph{runtime prompt} consists of an ordered access sequence followed by two JSON objects containing structured statistics and review context.
The three groups are labeled \texttt{Access sequence}, \texttt{Structured statistics}, and \texttt{Review context}, as in Figure~\ref{fig:brain-design}.
Table~\ref{tab:input-evidence-definitions} defines the structured statistics; the notation used in its formulas is specified below.

\clearpage
\begin{table}[!t]
\centering
\small
\setlength{\tabcolsep}{5pt}
\renewcommand{\arraystretch}{1.45}
\begin{tabularx}{\linewidth}{@{}>{\raggedright\arraybackslash}p{0.30\linewidth}>{\raggedright\arraybackslash}p{0.26\linewidth}>{\raggedright\arraybackslash}X@{}}
\toprule
\textbf{Field} & \textbf{Meaning} & \textbf{Definition} \\
\midrule
\multicolumn{3}{@{}l}{\textit{Scalar fields in both \texttt{window} and \texttt{recent\_tail}}} \\
\texttt{length} & Number of accesses & $n$ \\[4pt]
\texttt{unique\_ratio} & Distinct IDs per access & $m/n$ \\[4pt]
\texttt{access\_entropy} & Diversity of ID frequencies & $-\sum_{u\in U}p_u\log_2p_u\,/\,\log_2m$ if $m>1$; $0$ otherwise \\[4pt]
\texttt{gini} & Inequality of ID frequencies & $\displaystyle\frac{\sum_{j=1}^{m}(2j-m-1)c_{(j)}}{mn}$ \\[4pt]
\texttt{top\_5\_access} & Share of accesses to the five most frequent IDs & $\displaystyle\frac{\sum_{j=1}^{\min(5,m)}c_{[j]}}{n}$ \\[4pt]
\texttt{reuse\_ratio} & Share of accesses that repeat an ID & $r/n$ \\[4pt]
\texttt{short\_reuse\_ratio} & Share of repeats with short gaps & $\displaystyle\frac{1}{r}\sum_{i\in R}\mathbf{1}[d_i<10]$ \\[4pt]
\texttt{mean\_reuse\_distance} & Mean gap between repeated IDs & $\displaystyle\frac{1}{r}\sum_{i\in R}d_i$ if $r>0$; $n$ otherwise \\[4pt]
\midrule
\multicolumn{3}{@{}l}{\textit{Fields inside \texttt{adjacent\_steps} in either summary, for $n>1$}} \\
\texttt{zero\_ratio} & Consecutive repeated IDs & $q_0/(n-1)$ \\[4pt]
\texttt{pm1\_ratio} & Unit-step traversal & $(q_{-1}+q_1)/(n-1)$ \\[4pt]
\texttt{small\_jump\_ratio} & Small nonzero ID steps & $\displaystyle\frac{1}{n-1}\sum_{0<|\delta|\leq5}q_\delta$ \\[4pt]
\texttt{large\_jump\_ratio} & Large ID steps & $\displaystyle\frac{1}{n-1}\sum_{|\delta|>100}q_\delta$ \\[4pt]
\texttt{dominant\_steps} & Most frequent ID steps & Up to five pairs $(\delta,q_\delta)$ with largest $q_\delta$, ordered by decreasing count; ties follow first appearance. \\[4pt]
\quad\texttt{delta}, \texttt{count} & Step value and its frequency & The two members of each pair: $\delta$ and $q_\delta$. \\[4pt]
\bottomrule
\end{tabularx}
\caption{Structured statistics computed from the access sequence. Ratios with zero denominators are $0$ unless an explicit fallback is shown. Ratios, entropy, and Gini are rounded to three decimal places; mean reuse gap is rounded to one. Lengths, step differences, and counts are integers.}
\label{tab:input-evidence-definitions}
\end{table}

Within the statistics object, \texttt{window} and \texttt{recent\_tail} correspond to \texttt{full\_window\_statistics} and \texttt{tail\_window\_statistics}.

Let $W=(b_1,\ldots,b_N)$ denote \texttt{Access sequence}, and let
\[
 t=\min\{N,200,\max\{10,\lfloor N/2\rfloor\}\},
 \qquad T=(b_{N-t+1},\ldots,b_N).
\]
The \texttt{window} object summarizes $W$; \texttt{recent\_tail} summarizes $T$ with the same scalar and adjacent-step fields, computed within each sequence.
For either analyzed sequence $X=(x_1,\ldots,x_n)$, define
\[
 U=\{x_i:1\leq i\leq n\},\quad m=|U|,\quad
 c_u=\sum_{i=1}^{n}\mathbf{1}[x_i=u],\quad p_u=c_u/n.
\]
Write $c_{(1)}\leq\cdots\leq c_{(m)}$ for ascending counts and $c_{[1]}\geq\cdots\geq c_{[m]}$ for descending counts.
For repeated accesses, let
\[
 R=\{i:\exists j<i,\ x_j=x_i\},\quad r=|R|,\quad
 d_i=i-\max\{j<i:x_j=x_i\}\quad(i\in R).
\]
All reuse gaps count sequence positions within the analyzed window, rather than elapsed time or byte distance.
For adjacent steps, let $\Delta_i=x_{i+1}-x_i$ and $q_\delta=\sum_{i=1}^{n-1}\mathbf{1}[\Delta_i=\delta]$.

For review context, let $g$ denote the diagnosis trigger, $s$ the controller state when the snapshot is taken, and $p$ the previously committed pattern.
The state defaults to \texttt{default} when unavailable; the pattern is \texttt{null} when no valid previous pattern is available.
Let $\mathcal{P}=\{\texttt{sequential},\texttt{random},\texttt{skewed}\}$.
Table~\ref{tab:input-controller-definitions} defines the context fields; $t$ is the tail length defined above.

\begin{table}[!ht]
\centering
\small
\setlength{\tabcolsep}{5pt}
\renewcommand{\arraystretch}{1.5}
\begin{tabularx}{\linewidth}{@{}>{\raggedright\arraybackslash}p{0.34\linewidth}>{\raggedright\arraybackslash}p{0.24\linewidth}>{\raggedright\arraybackslash}X@{}}
\toprule
\textbf{Review-context field} & \textbf{Meaning} & \textbf{Value} \\
\midrule
\texttt{review\_trigger} & Reason for requesting diagnosis & $\texttt{chr\_anomaly}$ if $g=\texttt{chr}$; $\texttt{periodic\_retry}$ otherwise \\[4pt]
\texttt{previous\_control\_state} & Earlier controller state & $s$ \\[4pt]
\texttt{previous\_committed\_pattern} & Earlier committed pattern & $p\in\mathcal{P}\cup\{\texttt{null}\}$ \\[4pt]
\texttt{older\_prefix\_may\_be\_stale} & Possible outdated access prefix & $(g=\texttt{chr})\lor(s=\texttt{transition})$ \\[4pt]
\texttt{recent\_tail\_length} & Number of tail accesses & $t$ \\[4pt]
\bottomrule
\end{tabularx}
\caption{Review-context fields in \texttt{Review context}. The stale-prefix flag is a Boolean warning, not an observed phase boundary.}
\label{tab:input-controller-definitions}
\end{table}

\section{Sensitivity to the CHR Reflection Threshold}
\label{sec:chr-threshold-sensitivity}

A lower CHR deviation threshold makes feedback reflection more sensitive: it
can request more diagnoses and react sooner to pattern changes, at the cost of
additional diagnosis tokens.
To examine this tradeoff, we replay the ResNet-50 workflow W1 from
Section~\ref{sec:eval-e2e}. We set both
the CHR drop and increase thresholds relative to the previous EWMA to 0.40,
0.20, and 0.10, while holding the other settings fixed.

\begin{figure*}[t]
\centering
\includegraphics[width=\textwidth]{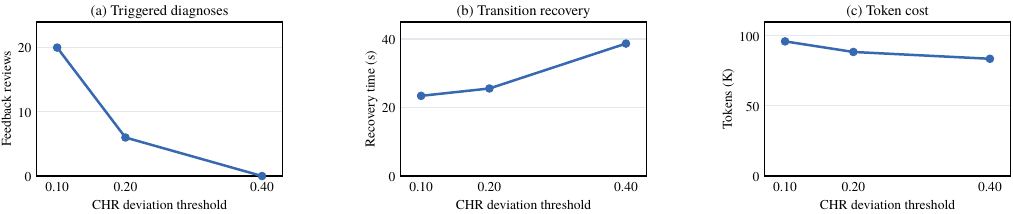}
\caption{Sensitivity to the CHR reflection threshold on the ResNet-50 workflow.
Both the drop and increase thresholds take the value shown. Recovery
time is the mean time from a phase boundary until an accepted
diagnosis commits the matching pattern.}
\label{fig:chr-threshold-sensitivity}
\end{figure*}

Figure~\ref{fig:chr-threshold-sensitivity} shows that lowering the threshold
from 0.40 to 0.10 increases feedback-triggered diagnoses from 0 to 20, while
mean recovery time across pattern transitions falls from 38.7 to
23.4~s. Total diagnosis tokens increase from 83.7k to 96.1k. The data indicates
that a more sensitive feedback gate can improve
adaptation at the expense of diagnosis cost. However, CHR remains nearly
identical across the settings, ranging from 69\% to 70\%. This is because,
as the dataset and its phases grow longer, the accesses made before a
periodic diagnosis corrects the pattern account for a smaller fraction of the
workflow. The resulting difference in cache hits is therefore diluted by the
many accesses served after the policy has adapted.

\end{document}